# Quantum echo-enabled high harmonic generation using ultrafast electrons

Yangfan He[1], Chenhao Pan[2], Bin Zhang[2,3], Yiming Pan[2*]

[1]School of Physics and Astronomy, China West Normal University, Nanchong 637002, China

[2]State Key Laboratory of Quantum Functional Materials, School of Physical Science and Technology and Center for Transformative Science, ShanghaiTech University, Shanghai 200031, China

[3]School of Electrical Engineering - Physical Electronics, Center of Laser-Matter Interaction, Tel Aviv University, Ramat Aviv 69978, Israel

**Abstract**

Controlling and generating ultrafast free-electron wavepackets via laser is pivotal for photon-induced near-field electron microscopes (PINEM) and also for developing compact, coherent free-electron radiation sources. Here, we present a quantum echo-enabled high-harmonic generation (QEEHG) scheme that manipulates the quantum phase of electron wavepackets to produce tunable, coherent high-harmonic radiation at ultrashort wavelengths. This framework, inspired by the EEHG concept for free-electron lasers by Stupikov et al. (2009), leverages multiphoton PINEM scattering followed by dispersive chirp sections to induce quantum interference among photon sidebands. Such interference selectively enhances a targeted harmonic order – for instance, the 60[th] harmonic at 13.3nm from an 800nm seeding - while suppressing unwanted radiations. The optimization of harmonic orders and its non-classical spectral characteristics are analyzed. This quantum echo technique establishes a promising paradigm for compact coherent sources and provides new perspectives for quantum wavefunction shaping in ultrafast electron microscopy and diffraction.

Coherent sources of extreme ultraviolet (EUV) light are indispensable tools[1] across multiple disciplines, enabling studies of ultrafast electron dynamics and supporting next-generation semiconductor manufacturing[2,3]. The development of such compact sources at EUV and X-ray wavelengths remains a central challenge in modern physics. Existing generation methods face severe limitations. Commercial EUV lithography machines produce 13.5nm radiation via laser-illuminated plasma from tin droplets, while lacks spatial and temporal coherence. Atomic high-harmonic generation (HHG) such as argon, which capable of producing coherent beams, suffers from intrinsically low efficiency, limited output power, and difficulties in phase matching and gas-phase stability[4,5]. Moreover, HHG necessarily generates all intermediate harmonic orders, resulting in inefficient use of energy conversion. Free-electron sources – synchrotrons and free-electron lasers (FELs)-provide a more scalable route to intense coherent emission. Modern FELs employ many radiative schemes such as self-amplified spontaneous emission (SASE)[6,7], high-gain harmonic generation (HGHG)[8,9], and echo-enabled harmonic generation (EEHG)[10-13]. In EEHG scheme, for instance, two laser modulators and two dispersive chicanes impose energy modulations that create a controlled phase-space "echo" feature, producing fine micro-bunching at nanometer and high-intensity radiation at a chosen high harmonic. Nevertheless, FEL facilities are extremely large and costly, limiting their availability and use.

Concurrently, the field of ultrafast transmission electron microscopy (UTEM) and free-electron quantum optics - founded on photon-induced near-field electron microscopy (PINEM)- are advancing rapidly[14-17]. A key insight is that PINEM interaction between free electrons and optical fields enables quantum electron radiations with characteristic surpassing their classical counterparts, including Quantum HHG[18], many body HHG[19], wavepacket-based Smith-Purcell radiation[20], quantum Cherenkov radiation[21], superradiant emission[22], and PINEM-mediated high-harmonic generation (PINEM-HHG)[23,24]. These radiative schemes originate from the preparation of quantum electron beams: PINEM process imprints an optical-phase-dependent modulation onto the electron wavefunction, generating a comb of discrete photon sidebands. Free-space propagation of these PINEM electrons then converts the energy modulation into a train of attosecond microbunches, a conversion robustly validated by recent experiments[25-29]. The similar conversion from energy to density modulation is also required in FELs such as EEHGs. This motivates us to pose a question: Can we engineer quantum interference among PINEM sidebands, using multi-stage laser modulation and free-drift control deliberately, in analogy to the EEHG procedure, to selectively enhance a single high-harmonic from sub-MeV electrons in the UTEM setup?

In this *Letter*, we present quantum echo-enabled high-harmonic generation (QEEHG), a mechanism that overcomes the limited harmonics selectivity and inefficient yield of the reported schemes such as PINEM-HHG[24]. This quantum echo protocol employs two sequential laser modulations of quantum free-electron wavepackets, separated by two drift sections, thereby realizing a multi-path quantum interferometer among PINEM sidebands. The first PINEM interaction prepares the electrons in a discrete spectrum dressed by photon sidebands, after which a highly-chirped drift imparts a sideband-dependent phase accumulation. A second PINEM scattering redistributes each of these sidebands into other sidebands, forming a quantum pathway network (as shown in Fig. 1b). The final drift enables a phase-sensitive recombination of all pathways to generate high harmonics from the echo-modulated electrons. We find that this echo technique achieves selective enhancement of targeted harmonic orders (e.g., EUV wavelengths of 13.3nm) while suppressing undesired harmonics through interference destruction among those pathways. Our work establishes QEEHG as a novel paradigm for laser control of ultrafast free-electron wavefunctions and generations of compact, coherent light sources.

***Setup and modelling.*** Our proposed experimental setup for QEEHG beamline is depicted in Fig. 1a, consisting of an ultrafast electron source, two optical modulation stages, two dispersive drifts (chirp1 and chirp2), and a downstream HHG radiation converter. In Modulator 1, electrons interact with a laser field at frequency $\omega$ via the near field of a metallic or dielectric nanostructure (e.g., a metallic nanotips array), producing coherent combs of photon sidebands in the electron spectrum through stimulated multiphoton exchange, this multiphoton scattering process distinct from classical energy modulation. Propagation through Chirp 1 (free space) then imprints a sideband-dependent phase due to longitudinal dispersion; for sub-MeV electrons the required dispersion can be provided by a millimeter-scale drift, offering a major simplification for experiments (see SM file for parameter range). The highly-chirped electron wavefunction is further processed in Modulator 2 by a second laser at frequency $\omega_2 = \eta\omega$ ($\eta \neq 1$ corresponds to two color operation), followed by additional drift during Chirp 2 before emitting short-wavelength radiations through the final HHG Converter.

The radiation from the above quantum-echo-modulated electron wavepacket is analyzed thorough the bunching factor [22], which quantifies Fourier components of the free-electron current operator observable responsible for HHG emission. The $q$-th harmonic bunching factor is $b^{(q)} = \langle \psi_f | \hat{b}^q | \psi_f \rangle$, with $\hat{b}^q = e^{iqk\hat{z}}$ is the bunching operator of the q-th order, where $\hat{z}$ is the propagation direction and $k = \omega/v_0$ denotes the longitudinal recoil wavevector associated with an energy exchange $\hbar\omega$ for an electron at velocity $v_0$. In a given converter, the spectral amplitude at yield $q\omega$ and is proportional to $b^{(q)}$ up to a converter dependent response and is applicable to free-electron radiative schemes of both classical beam modulation and quantum wavefunction shaping[20,23,24]. The final electron wavefunction $|\psi_f\rangle$ is derived from the initial state $|\psi_e^0\rangle$ using the above quantum echo process given by

$$|\psi_f\rangle = \hat{U}_{\text{free}}(t_2)\hat{M}(g_2)\hat{U}_{\text{free}}(t_1)\hat{M}(g_1)|\psi_e^0\rangle \qquad (1)$$

where each laser modulation with nanotips is explicitly obtained $\hat{M}(g_i) = \exp(-ig_i \sin(k_i\hat{z} + \phi_i))$ ( $i = 1,2$ )[17], with $k_i = \omega_i/v_0$, $g_i$ the dimensionless modulation strength proportional to the optical field strength within short-time interaction limit, and $\phi_i$ the laser phase relative to the modulated electron. Free space propagation over a drift time $t$ is describe by the chirping operator is $\hat{U}_{\text{free}}(t) = \exp(-i\xi t\hat{k}^2)$, where $\xi$ is the chirping coefficient; experimentally $t$ is tuned by adjusting the drift lengths $d_1$ and $d_2$. We ignore the chirping effect during the modulations because the PINEM interaction time is much shorter than the free-drift times. Full derivations of the quantum echo process from time-dependent Schrodinger equation can be found in Section 1 of the Supplementary Material.

Applying the sequence of operators in Eq. (1) leads to our central result: an explicit formula for the intensity of the $q$-th harmonic bunching factor

$$|b^{(q)}|^2 = \left|\sum_{q_2} e^{iq_2\Theta} J_{q-\eta q_2}\left(2g_1 \sin\left(\mathcal{M}_1^{(q_2)}\right)\right) J_{q_2}(2g_2 \sin(\mathcal{M}_2))\mathcal{U}^{(q_2)}\right|^2 \qquad (2)$$

This explicit expression describes the full process of quantum echo-enabled high harmonic generation. The summation runs over all possible intermediate quantum pathways indexed by $q_2$, the net number of photons absorbed or emitted in the second modulator. The term $\Theta = \eta\phi_1 - \phi_2 + \frac{\pi}{2}(\eta - 1)$ is the phase difference between two lasers. The path-dependent phase is $\mathcal{M}_1^{(q_2)} = ((q - \eta q_2)\xi t_1 + q\xi t_2)k_1^2$, the common phase for all paths is $\mathcal{M}_2 = \eta q\xi t_2 k_1^2$. The envelope function $\mathcal{U}^{(q_2)} = \exp\left(-2((q\xi(t_1 + t_2) - \eta q_2\xi t_1)\sigma_k k_1)^2\right)$ accounts for the contribution due to the intrinsic momentum spread $\sigma_k$. The Bessel functions $J_{q-\eta q_2}$ and $J_{q_2}$ encode the probability amplitudes for multiphoton processes in two modulators, respectively. Full derivations and discussions of these terms are provided in Section 2 of the Supplementary Material.

The case $\eta = 1$, corresponding to identical modulator frequencies, is experimentally favorable: it enables single-source optical driving, facilitates precise phase locking between modulation stages, and serves as a model system for quantum echo control with commensurate-frequency lasers (see the following section). For non-integer $\eta$, Eq. (1) remain valid, but $b^{(q)}$ is evaluated numerically.

***QEEHG scheme.*** Equation 2 makes explicitly that QEEHG results from the coherent superposition of multiphoton scattering channels. Therefore, we schematically formalize the QEEHG process as a multi-path quantum network (Fig. 1(b)), with each element of the beamline implementing a well-defined quantum operation:

The initial wavefunction $|\psi_e^0\rangle$ emitted from a laser-excited electron gun defines the network's starting point. Modulator 1 at frequency $\omega_1$, prepares a coherent superposition of photon sidebands ($\pm n\hbar\omega_1$), with population set by the modulation strength $g_1$. Chirp 1 imprints a pathway dependent phase across each sideband. This process converts the discrete energy modulation to highly-chirped modulated probability density distribution. Modulator 2 at frequency $\omega_2 = \eta\omega_1$ further branches each populated sideband into additional channels with amplitudes controlled by $g_2$. For instance, sideband $|n\hbar\omega_1\rangle$, scatters into new sideband $|n\hbar\omega_1 \pm m\eta\hbar\omega_1\rangle$ and forming a network with energy-quantized by $\eta\hbar\omega_1$ and scattering magnitude weighted by $g_2$ of the second modulator. Chirp 2 performs a second dispersive phase processing that enables controlled recombination of all pathways. Finally, the measurement harmonic bunching $|b^{(q)}|$ quantifies the net interference outcome of this network and provides the electrons-state from factor that drives radiation at $q\omega_1$ in a given converter.

The function of each beamline element is clarified by turning off individual operations in Eq. (1) and examining the results $b^{(q)}$(the detailed derivation can be found in SM Section 3). For instance, without initial sideband preparation ($g_1 = 0$), Chirp 1 acts trivially and Eq. 2 reduces to the single-stage PINEM-HHG[24] limit, $|b^{(q)}|^2 = \left|J_{\frac{q}{\eta}}(2g_2 \sin(\eta q \xi t_2 k_1^2))\exp(-2(q\xi t_2 \sigma_k k_1)^2)\right|^2$, governed solely by Modulator 2, with intrinsic-spread envelope proportional to $\exp(-2(q\xi t_2 \sigma_k k_1)^2)$. Removal of Chirp 1 ($t_1 = 0$), the sideband-dependent phase is absent and two modulators act effectively as a single interaction with a two-color field, yielding $|b^{(q)}|^2 =$ $|\exp(-2(qt_2\sigma_k k_1)^2) J_q(2(g_1 \sin(qt_2 k_1^2) + g_2 \sin(\eta q t_2 k_1^2)))|^2$, a two-color PINEM HHG response without pathway selectivity. On the other hand, If Modulator 2 is off ($g_2 = 0$), the bunching factor reduces to a single-modulator architecture $|b^{(q)}| = |J_q(2g_1 \sin(q\xi(t_1+t_2) k_1^2)) e^{-2(q\xi(t_1+t_2)\sigma_k k_1)^2}|^2$, controlled by $g_1$ and total drift time $t_1 + t_2$. Besides, if Chirp 2 is suppressed ($t_2 = 0$), the pathway sum collapses to $q_2 = 0$, the $q^{th}$ bunching becomes independent of $g_2$ $|b^{(q)}| = |J_q(2g_1 \sin(q\xi t_1 k_1^2)) e^{-2(q\xi t_1 \sigma_k k_1)^2}|$. This highlight that, Chirp 2 is the active phase - processing stage that enables the second modulation to participate in controlling sidebands interference. Collectively, these limits show that the chirp sections are not passive drifts but essential dispersive phase processors that make the pathway interference underlying QEEHG.

The QEEHG framework produces a highly selective bunching response, producing narrow peaks in $|b^{(q)}|$ at discrete "echo" conditions, as shown in Fig. 2a. For the parameters set of Fig. 2f, displaying a pronounced single harmonic at the first echo order of $q = 60$. The analytic predictions from Eq. (2) (blue bars) and numerical simulations (orange dots) are in excellent quantitative agreement, validating the full quantum echo framework. The corresponding Wigner distribution in Fig. 2b captures an intuitive view the free-electron wavepacket dynamics: the discrete sidebands reflect coherent multiphoton exchange, while the observed stretched interference fringes and fine phase-space structures visualize how dispersive drift and subsequent modulation create phase correlations required for pathway interference.

Figure 2c plots the envelope $\mathcal{U}^{(q_2)}$ as a function of the harmonic order $q$ and pathway index $q_2$, demonstrating that in the intrinsic momentum spread strongly suppresses most intermediate pathway so that only a limited pathway contributes to $b^{(q)}$. This filtering, together with the pathway-dependent phases in Eq. (2), indicates the observed selectivity is a purely quantum effect. Consistently, the channel-resolved analyses in Figs. 2e and 2f further decompose the contributions for $q = 59$ and $q = 60$, respectively, reveal constructive (red) and destructive (green) interference across sidebands. Confirming that quantum echo enables active control enhancing a target harmonic while suppressing neighbouring channels. The simulation methods are detailed in Supplementary Section 4.

The non-classical validation of the QEEHG scheme is fundamentally tied to its sensitivity to the ordering of the modulations and dispersive drifts in Eq. 1. Because the modulation operators depend on position ($\hat{z}$), while chirping operators depend on the conjugate momentum operators ($\hat{k}$) with $[\hat{z}, \hat{k}] = i$, $\widehat{\mathcal{M}}(\hat{z})$ and $\widehat{U}_{free}(\hat{k})$ do not commute, and interchanging their order changes the resulting $b^{(q)}$. In phase space terms: Modulator 1 imprints a spatially periodic phase, creating discrete momentum-space sidebands; the Chirp 1 introduces a dispersive phase that maps momentum into real-space phase structure; Modulator 2 scatters these free-electron states into additional channels; and the Chirp 2 drives final dispersion that allows pathway coherent recombination. The selective enhancement of qth bunching thus arise from an ordering dependent chain that engineers constructive interference at chosen order q.

***Quantum echo optimization.*** The qth harmonic spectrum $|b^{(q)}|^2$ depends on six-dimensional control parameters $\Pi_{\eta,\lambda_1} = \{|g_1|, \phi_1, d_1, |g_2|, \phi_2, d_2,\}$, comprising the modulation strengths, optical phases, and drift length. Our goal is to maximize the bunching factor for a chosen harmonic

$$\Pi^*_{\eta,\lambda_1} = \arg\max_{\Pi_{\eta,\lambda_1}} |b^{(q_d)}|, \tag{3}$$

which formally defines QEEHG as a quantum control problem. Physically, this amounts to tuning the modulation strengths, chirp lengths, and laser phases so that all contributing pathways interfere constructively at $q$th order and destructively elsewhere. Under ideal conditions, this reduces to

$$\text{Target: } \max|b^{(q_d)}|, \text{Constraint: } |b^{(q)}| \to 0 \text{ for } 1 \leq q < q_d \tag{4}$$

This framework generalizes naturally to higher-order modulation–dispersion sequences, where Eq. 2 must be evaluated numerically but Eq. 3 remains the defining optimization criterion. Mathematically, the task is equivalent to identifying a hyperplane in parameter space that renders the target pathway fully coherent and suppresses non-target contributions via destructive interference.

The chirping process $\mathcal{M}_2$ acts as a single global control knob for the entire sideband interference network. In practice, tuning the second drift length $d_2$ synchronously accumulates all pathway phases, providing a straightforward handle for coarse optimization. As illustrated in Fig. 3, we implement a two-stage control protocol to obtain the final parameter optimalization. We first randomly choose other parameters, sweep $d_2$ to meet the condition $|2g_2 \sin(\mathcal{M}_2)| \pm \left(q + q^{\frac{1}{3}}\right) \approx 0$, which establishes constructive interference across all contributing pathways. Next, the first modulation strength $|g_1|$ and drift $d_1$ are subsequently fine-tuned to adjust $\mathcal{M}_1^{(q_2)}$ and optimize relative phase $(\phi_1 - \phi_2)$, acting as a spectral filter that enhance targeted harmonics. Based on this optimization procedure, we find deterministic engineering of the harmonic spectrum, producing periodic enhancement of selected orders (e.g., every 13th harmonic) and distinct step-like plateaus, as shown in Figs. 3b–3c.

***Further Discussions.*** Achieving high-power radiation requires dense electron bunches, but space-charge forces for ultrafast electron beams can degrade the quantum coherence needed for QEEHG. For sub-MeV electrons, Coulomb repulsion would significantly limit the harmonic radiation performance. We address that scaling electron sources to $10^{7-9}$ electrons per pulse while preserving coherence remains a key hurdle for practical realization of multiparticle quantum echo.

Recent advances in nanofabrication offer a promising route. Metallic nanotip arrays, fabricated via electron-beam lithography, allow tip sizes below ~3 nm and inter-tip spacing near ~100 nm. These arrays enable large-scale gated field emitters with $10^4$–$10^6$ tips, producing up to 5 pC bunch charge within 50 fs pulses. The resulting electron beams show dramatically enhanced transverse coherence and reduced emittance—critical for mitigating space-charge limitations. Experimental validation of this approach was provided by Tsujino et al.[30-32], with their demonstration of electron bunches reaching $10^7$ electrons generated from a $1.2 \times 10^5$ molybdenum tip array irradiated by 50 fs laser pulses under a 100 V bias, corresponding to an average of 100 electrons per tip per pulse[33-35]. Such performance approaches the charge densities required for compact X-ray FELs, underscoring the feasibility of implementing QEEHG at practically relevant beam currents.

**Conclusion.** We have proposed and analyzed quantum echo-enabled high-harmonic generation (QEEHG) as a framework for selective harmonic emission from laser-modulated free-electron wavepackets. Our combined analytical and numerical treatment reveals three regimes of spectral response – single harmonic order, periodically echoed harmonics, and platform-like – arising from constructive sideband interference among quantum pathways. QEEHG fundamentally departs from classical harmonic-generation schemes by exploiting cumulative phase evolution under laser modulation and free drift, thereby coherently programming and optimizing the electron wavefunction shaping. This approach provides a promising pathway to compact, coherent short-wavelength sources in EUV regimes and offers a versatile platform for quantum coherent control of free-electron systems with potential applications in ultrafast electron microscopy and imaging.

**Acknowledgement.**

We thank Senlin Huang for insightful discussions. We would like to acknowledge the support of the NSFC (No. 2023X0201-417-03), the fund of the ShanghaiTech University (Start-up funding), the Sichuan Science and Technology Program (No.2025ZNSFSC0855) and the Research Funds of China West Normal University (No. 23KE025).

Correspondence and requests for materials should be addressed to Y.P.: yiming.pan@shanghaitech.edu.cn.

**Data and code availability**

The code repository will be made publicly available on acceptance through an open-source link.

**Conflict of interest**

The authors declare no competing interests.

**Figures:**

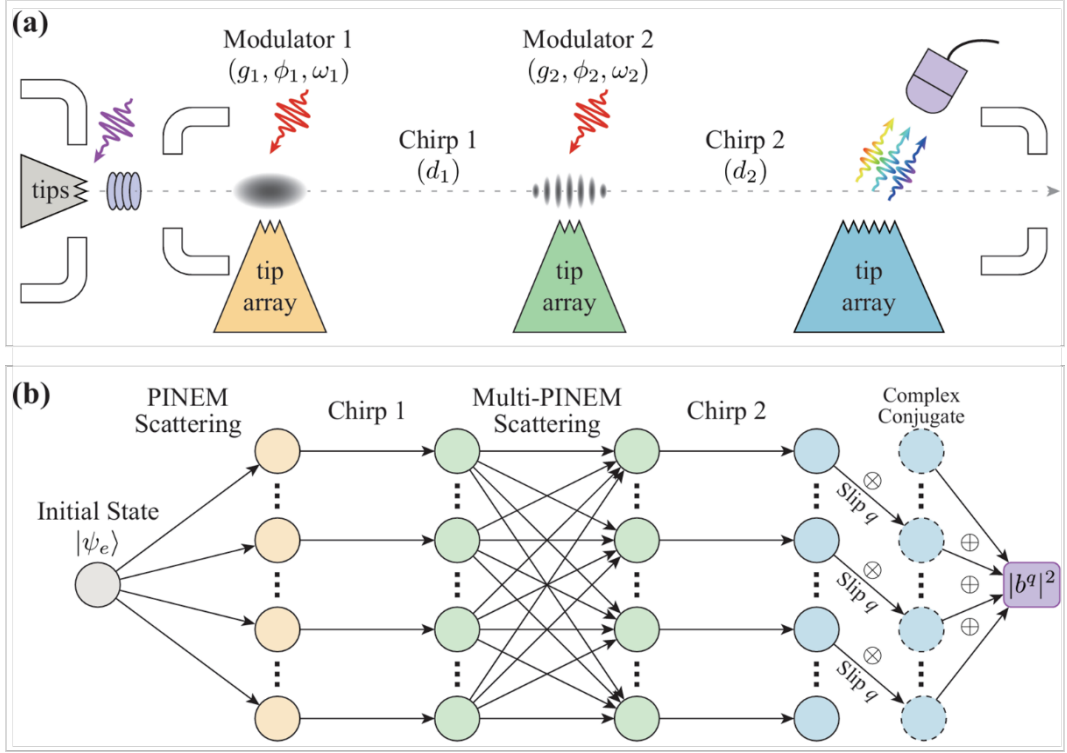

**Figure 1: The setup and quantum process of Echo-Enabled High-Harmonic Generation (QEEHG).** (a)The schematic setup of quantum echo-enabled high harmonic generation (EEHG) using free electrons. The setup is based on the ultrafast transmission electron microscopy that resembles a EEHG free electron laser seeding technique, which involves the usages of two laser modulations (modulator #1 and #2) and twice long-time free propagations (chirp #1 and chirp #2) to create a high-harmonic electron density modulation, in order to produce the high harmonics at the desired order by suppressed the other harmonic generations. (b)Multi-Path Quantum Interference Network representation of the QEEHG process. It consists of six parts: (1) Initial electron wavepacket; (2) First modulation interaction creating multiple energy sidebands; (3) First chirp imprinting momentum-dependent phases; (4) Second modulation interaction generating parallel multi-PINEM channels; (5) Second chirp enabling coherent pathway interference; (6) Final harmonic spectrum $|b^{(q)}|$.

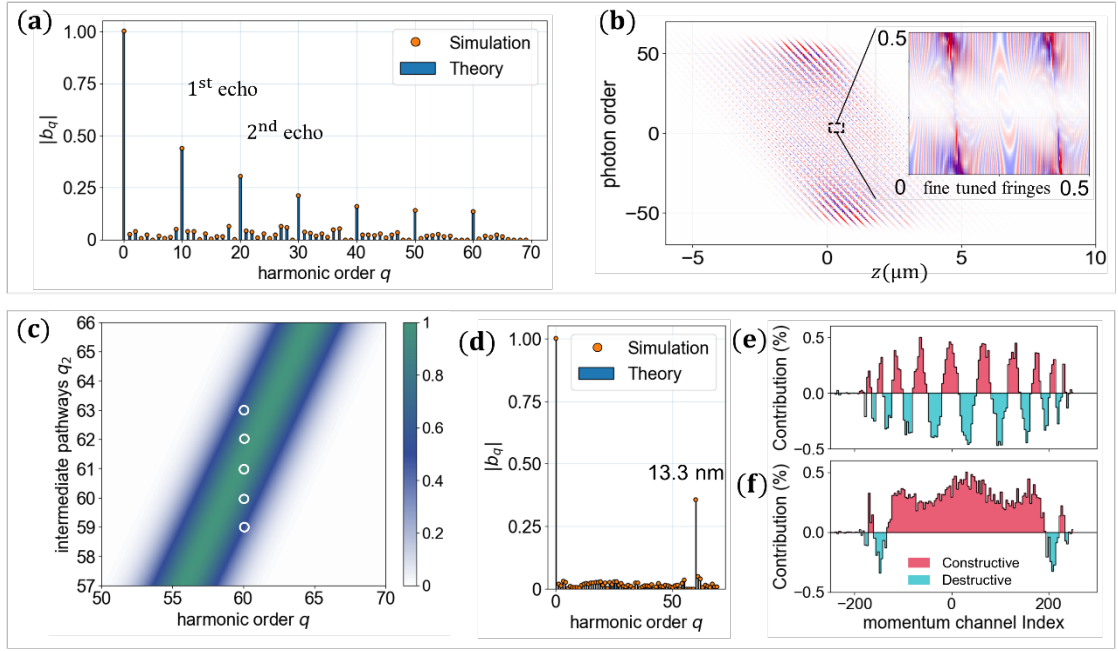

**Figure 2: Quantum control and path selectivity in QEEHG.** (a) Harmonic spectrum showing multiple echo orders. Blue bars: theoretical prediction from Eq. (3). Orange dots: numerical simulation results. (b) Wigner distribution corresponding to (a), showing quantum coherence signatures. (c) Path selection map $\mathcal{U}^{(q_2)}$ vs. $q$ and $q_2$. Black circles denote dominant coherent pathways. (d) Highly selective harmonic generation at $q = 60$. (e,f) Contributions of different quantum pathways to the (e) $q = 59$ and (f) $q = 60$ harmonics, showing constructive (green) and destructive (grey) interference. Parameters for Figs. 3(a-b): $\Pi^*_{1,0.8\mu m} = \{5, 244\text{mm}, 0, 60, 25.8\text{mm}, 0\}$, $\Pi^*_{1,0.8\mu m} = \{2, 210\text{mm}, 0, 240, 4.34\text{mm}, 0\}$ for (c-f).

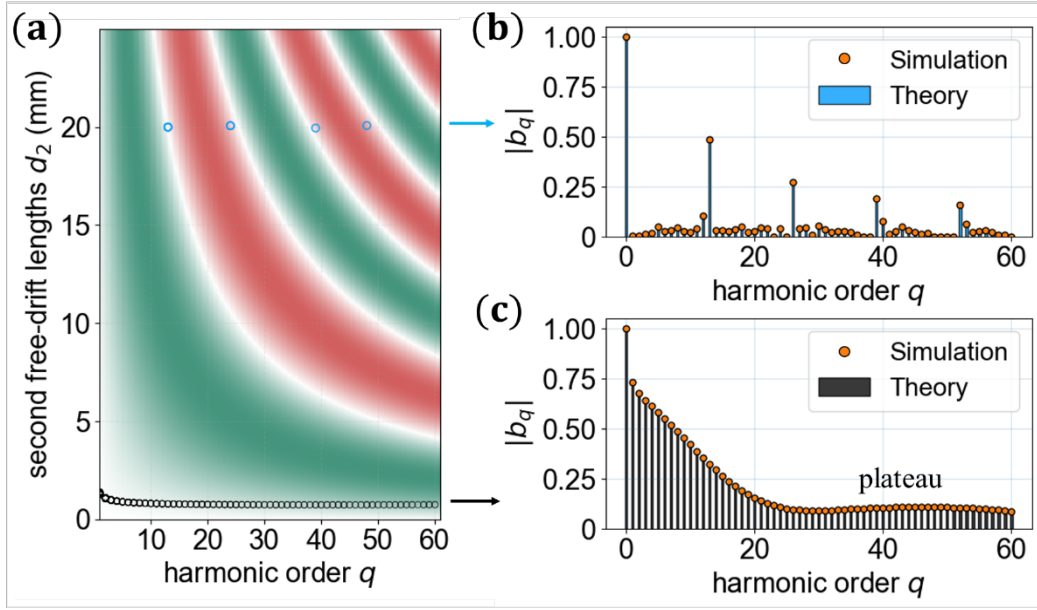

**Figure 3: Tailored harmonic spectroscopy through collaborative phase optimization.** (a) Identifying optimal global phase conditions by scanning the second drift length ($d_2$). Black and blue circles mark regions satisfying the condition $|2g_2 \sin(\mathcal{M}_2)| \pm \left(q + q^{\frac{1}{3}}\right) \approx 0$. (b, c) Subsequent pathway selection via fine-tuning of the first modulator's parameters ($g^1, \phi_1, d_1$), yielding spectrally tailored outputs: (b) periodic enhancement (every 13th harmonic) and (c) step-like (plateau).

# Supplemental Material：

# Quantum echo-enabled high harmonic generation using ultrafast electrons

Yangfan He[1], Chenhao Pan[2], Bin Zhang[2,3], Yiming Pan[2]

[1]School of Physics and Astronomy, China West Normal University, Nanchong 637002, China

[2]State Key Laboratory of Quantum Functional Materials, School of Physical Science and Technology and Center for Transformative Science, ShanghaiTech University, Shanghai 200031, China

[3]School of Electrical Engineering - Physical Electronics, Center of Laser-Matter Interaction, Tel Aviv University, Ramat Aviv 69978, Israel

## 1. Laser modulation and chirping operators

To elucidate the physical origin of the laser modulation and chirping operators used in the main text and to verify the quantum echo evolution described in Eq. (1), we begin with the time-dependent Schrödinger equation for a free electron interacting with an optical field:

$$i\hbar \frac{\partial}{\partial t}\psi(z,t) = \widehat{H}\psi(z,t) = (\widehat{H}_0 + \widehat{H}_I)\psi(z,t) \qquad (S1)$$

Here, the kinetic Hamiltonian $\widehat{H}_0$ for a relativistic free electron is expressed as:

$$\widehat{H}_0 = E_0 + v_0(\hat{p} - p_0) + \frac{(\hat{p} - p_0)^2}{2\gamma^3 m_e} \qquad (S2)$$

where $E_0$ and $p_0 = \gamma m_e v_0$ denote the central energy and momentum of the electron wavepacket, $m_e$ is the electron rest mass, and $\gamma$ is the Lorentz factor. We choose the initial kinetic energy $E_0$=200keV in the main text.

The interaction Hamiltonian $\widehat{H}_I$, under the Coulomb gauge and within the co-propagating laser field approximation, takes the form:

$$\widehat{H}_I = -\frac{e}{\gamma m} A(z,t) \cdot \hat{p} + \frac{i\hbar e}{2\gamma m_e} \frac{\partial A}{\partial z} \qquad (S3)$$

This interaction accounts for momentum exchange between the electron and a laser pulse of frequency $\omega_L$.

Having established the full Hamiltonian in Eq. (S1), we can solve the time dependent Schrodinger equation by employing the interaction picture. The evolution operator $\hat{S}$ in this picture with the interaction duration $t_{\text{int}}$ is:

$$\hat{S} = \mathcal{T}\exp\left(-\frac{i}{\hbar} \int_0^{t_{\text{int}}} \widehat{H}_I^{(I)}(t')\, dt'\right) \qquad (S4)$$

where $\mathcal{T}$ is the time-ordering operator, and $\widehat{H}_I^{(I)}(t) = e^{i\widehat{H}_0 t/\hbar} \widehat{H}_I e^{-i\widehat{H}_0 t/\hbar}$ is the interaction Hamiltonian in the interaction picture.

For a laser field with a longitudinal electric component $E(z,t) = E_1 \cos(\omega_1 t - k_1 z + \phi)$, the associated vector potential is $A(z,t) = -(E_1/\omega_1)\sin(\omega_1 t - k_1 z + \phi)$. We now consider the transformation of the operators in $\hat{H}_I^{(I)}(t)$. Under the free evolution governed by $\hat{H}_0$, and considering the electron's central momentum $p_0$, the position operator evolves approximately as $\hat{z}(t) \approx \hat{z} + v_0 t$, where $\hat{z}$ is the Schrödinger-picture operator. Substituting this into the phase of the vector potential, we obtain:

$$\omega_1 t - k_1 \hat{z}(t) + \phi \approx \omega_1 t - k_1(\hat{z} + v_0 t) + \phi = -k_1 \hat{z} + (\omega_1 - k_1 v_0)t + \phi \quad (S5)$$

For a laser co-propagating with the electron, the wavevector $k_1 = \omega_1/v_0$, leading to the resonance condition $\omega_1 - k_1 v_0 = 0$. Consequently, the explicit time dependence in the phase is suppressed, and the phase reduces to $-k_1 \hat{z} + \phi$.

Substituting this phase into the interaction Hamiltonian and noting that the dominant contribution comes from the term proportional to $\hat{p}A(z,t)$, the integral in Eq. (S4) can be evaluated. Under the assumption of an impulsive, short interaction ($t_{\text{int}} \to 0$), the time-ordering can be neglected. The scattering operator then becomes:

$$\hat{S} \approx \exp\left(-\frac{i}{\hbar}\int \hat{H}_I(t')\,dt'\right) \approx \exp\left(-i\frac{eE_1 v_0 t_{\text{int}}}{\hbar\omega_1}\sin(k_1\hat{z} + \phi)\right) \quad (S6)$$

This can be treated as the **laser modulation operator** which is used in the main text:

$$\hat{M}(g_i) = \exp(-ig_i \sin(k_i\hat{z} + \phi_i)) \quad (S7)$$

with the dimensionless modulation strength $g$ is approximately defined as:

$$g_i = \frac{eE_i v_0 t_{\text{int}}}{\hbar\omega_i} \quad (S8)$$

Following each laser modulation steps, the electron wavepacket undergoes a free propagation distance in vacuum, governed solely by the kinetic Hamiltonian $\hat{H}_0$ from Eq. (S2). The objective here is to derive the operator that describes this evolution and connect it to the form used in the main text.

Meanwhile, the evolution operator over a duration $t$ for free propagation is given by

$$\widehat{U}_{\text{free}}(t) = \exp\left(-\frac{i}{\hbar}\widehat{H}_0 t\right) \quad (S9)$$

To understand the effect of this operator on the electron's wavefunction, it is most insightful to consider the dynamics in a frame co-moving with the electron's central momentum $p_0$. In the comoving frame, the first two terms of $\widehat{H}_0$, $E_0 + v_0(\hat{p} - p_0)$, contribute to a global phase and a coordinate translation, respectively, which do not affect the internal structure of the wavepacket.

The dominant effect comes from the dispersive term, $\frac{(\hat{p}-p_0)^2}{2\gamma^3 m_e}$. Expressing the momentum deviation in terms of the wavevector, $\hat{k} = (\hat{p} - p_0)/\hbar$, this key term becomes $\frac{\hbar^2 \hat{k}^2}{2\gamma^3 m_e}$. Substituting this back into the evolution operator, and neglecting the overall phase and translation terms for clarity, we obtain the **chirping operator** that governs the dispersive dynamics:

$$\widehat{U}_{\text{free}}(t) = \exp\left(-\frac{i}{\hbar} \cdot \frac{\hbar^2 \hat{k}^2}{2\gamma^3 m_e} \cdot t\right) \quad (S10)$$

This is exactly the form presented in the main text:

$$\widehat{U}_{\text{free}}(t) = \exp(-i\xi t \hat{k}^2) \quad (S11)$$

where the chirping coefficient is defined as $\xi = \hbar/(2\gamma^3 m_e)$. Experimentally, the chirping duration $t = d/v_0$ can be controlled by the free-drift lengths $d_1$ and $d_2$.

**2. Harmonic bunching factor**

**2.1. Quantum Echo State Evolution**

In the main text, the harmonic bunching factor $b^{(q)} = \langle \psi_f | \hat{b}^q | \psi_f \rangle$ is identified as the key quantity governing high-harmonic generation. To compute $b^{(q)}$, we must first derive the explicit form of the final electron state $|\psi_f\rangle$ resulting from the quantum echo sequence. This section details the derivation of $|\psi_f\rangle$, starting from an initial

Gaussian wavepacket and sequentially applying the modulation and free-evolution operators derived in Section 1.

The protocol begins with an input quantum electron wave (QEW) state characterized by a Gaussian envelope in the wave-vector (k) basis:

$$|\psi_e^0\rangle = \sum_k c_k |k\rangle \tag{S12}$$

where $c_k = \frac{1}{\sqrt{2\pi\sigma_k}} \exp\left(-\frac{k^2}{4\sigma_k^2}\right)$. The first operation is phase modulation produced by a laser beam of dimensionless modulation depth $g_1$, wave-vector $\delta k$, and phase $\phi_1$. Applying $\widehat{M}(g_1)$ to the initial state yields

$$|\psi_e^1\rangle = \widehat{M}(g_1)|\psi_e^0\rangle = e^{-ig_1 \sin(k_1\hat{z}+\phi_1)}|\psi_e^0\rangle \tag{S13}$$

Immediately after the first modulation the QEW propagates freely for a time $t_1$. And the state after first chirping is

$$|\psi_e^2\rangle = \widehat{U}_{\text{free}}(t_1)\widehat{M}(g_1)|\psi_e^0\rangle = e^{-i\xi t_1 \hat{k}^2} e^{-ig_1 \sin(k_1\hat{z}+\phi_1)}|\psi_e^0\rangle. \tag{S14}$$

A second modulation is now applied by a second laser whose frequency is $\eta\omega_1$. Its wave-vector is therefore $\eta k_1$, its dimensionless modulation depth $g_2$, and its phase $\phi_2$. Acting with $\widehat{M}(g_2)$ on $|\psi_e^2\rangle$ gives, we obtain

$$|\psi_e^3\rangle = \widehat{M}(g_2)\widehat{U}_{\text{free}}(t_1)\widehat{M}(g_1)|\psi_e^0\rangle = e^{-ig_2 \sin(\eta k_1\hat{z}+\phi_2)}|\psi_e^2\rangle \tag{S15}$$

Finally, the QEW is allowed to propagate freely once more for a time $t_2$. Hence, the final state $|\psi_f\rangle$ after the second bunching is

$$|\psi_f\rangle = \widehat{U}_{\text{free}}(t_2)\widehat{M}(g_2)\widehat{U}_{\text{free}}(t_1)\widehat{M}(g_1)|\psi_e^0\rangle$$
$$= e^{-i\xi t_2 \hat{k}^2} e^{-ig_2 \sin(\eta k_1\hat{z}+\phi_2)} e^{-i\xi t_1 \hat{k}^2} e^{-ig_1 \sin(k_1\hat{z}+\phi_1)}|\psi_e^0\rangle. \tag{S16}$$

By applying the Jacobi–Anger expansion:

$$e^{-ig\sin(k\hat{z}+\phi)} = \sum_n J_n(g)e^{-in\phi}e^{-ink\hat{z}}, \tag{S17}$$

where $J_n$ is the nth-order Bessel function of the first kind.

We define $\theta_1 = k_1\hat{z} + \phi_1$ and $\theta_2 = \eta k_1\hat{z} + \phi_2$, inserting Eq. (S17) into Eq. (S16) and pulling the Bessel-function prefactors in front of the operators gives:

$$|\psi_f\rangle = \sum_n J_n(g_2)e^{-in\phi_2}\sum_m J_m(g_1)\,e^{-im\phi_1}e^{-i\xi t_2\hat{k}^2}e^{-in\eta k_1\hat{z}}e^{-i\xi t_1\hat{k}^2}e^{-imk_1\hat{z}}|\psi_e^0\rangle. \tag{S18}$$

This side-band expansion is crucial as it replaces the intractable sinusoidal operators with exponentials of the position operator, which are straightforward momentum displacement operators.

The next step is to evaluate the action of the alternating sequence of displacement and chirping operators on the initial state. Since the displacement operator $e^{ik_1\hat{z}}$ and the chirping operator $e^{-i\xi t\hat{k}^2}$ do not commute, an appropriate algebraic identity is required to combine them.

**2.2. Operator Reordering and the Final State Expression**

In Eq. (S18), the final state is expressed as a sum over terms containing alternating sequences of displacement operators $e^{ik_1\hat{z}}$ and chirping operators $e^{-i\xi t\hat{k}^2}$. To evaluate the action of this operator sequence on the initial state $|\psi_e^0\rangle$, we need a method to simplify these non-commuting operator products. Our goal is to derive an algebraic identity that allows us to "reorder" these operators, specifically by moving all displacement operators to one side, which will greatly simplify the subsequent calculation.

To establish the relation between the operator product $e^{-i\xi t\hat{k}^2}e^{ik_1\hat{z}}$ and a product with the reversed order, we analyze their action on the complete basis of momentum eigenstates $\{|k\rangle\}$. Beginning with the action on an arbitrary state $|k\rangle$, we first apply the displacement operator, which acts as a translation in momentum space, yielding $e^{ik_1\hat{z}}|k\rangle = |k+k_1\rangle$. Proceeding to apply the chirping operator, which imparts a phase proportional to the square of the momentum, gives $e^{-i\xi t\hat{k}^2}|k+k_1\rangle = e^{-i\xi t(k+k_1)^2}|k+k_1\rangle$. Consequently, the combined action of the product is $e^{-i\xi t\hat{k}^2}e^{ik_1\hat{z}}|k\rangle = e^{-i\xi t(k+k_1)^2}|k+k_1\rangle$.

To obtain an equivalent expression with the operators reordered, we consider a candidate operator of the form $e^{ik_1\hat{z}}e^{-i\xi t(\hat{k}+a)^2}$ and determine the constant $a$ required to reproduce the same result. Acting with the modified chirping operator first, we find $e^{-i\xi t(\hat{k}+a)^2}|k\rangle = e^{-i\xi t(k+a)^2}|k\rangle$. Subsequently applying the displacement operator gives $e^{ik_1\hat{z}}(e^{-i\xi t(k+a)^2}|k\rangle) = e^{-i\xi t(k+a)^2}|k+k_1\rangle$. For this final result to be identical to the previous expression for all $|k\rangle$, the condition $(k+a)^2 = (k+k_1)^2$ must hold, which is satisfied if and only if $a = k_1$. Therefore, we have shown that for any momentum eigenstate $|k\rangle$: $e^{-i\xi t\hat{k}^2}e^{ik_1\hat{z}}|k\rangle = e^{ik_1\hat{z}}e^{-i\xi t(\hat{k}+k_1)^2}|k\rangle$. Since the momentum eigenstates form a complete basis, the operators acting on them must be identical. This establishes the key commutation identity:

$$e^{-i\xi t\hat{k}^2}e^{ik_1\hat{z}} = e^{ik_1\hat{z}}e^{-i\xi t(\hat{k}+k_1)^2}. \qquad (S19)$$

This identity is the crucial tool we need. It allows us to systematically move all displacement operators in Eq. (S18) to the left of the chirping operators, paving the way for a direct evaluation of the final wavefunction $|\psi_f\rangle$ in the subsequent section.

Exploiting the reordering identity derived in Eq. (S19), we can systematically commute all displacement operators to the left of the chirping operators in the expression for the final state. Focusing on the operator sequence for a specific (n,m) side band pair in Eq. (S18), $e^{-i\xi t_2 \hat{k}^2} e^{-in\eta k_1 \hat{z}} e^{-i\xi t_1 \hat{k}^2} e^{-imk_1 \hat{z}}$, our objective is to resolve the non-commutativity between the chirps and displacements.

We begin by applying the identity of Eq. (S19) to the innermost pair of operators, $e^{-i\xi t_1 \hat{k}^2} e^{-imk_1 \hat{z}}$. The identity of Eq. (S19) allows us to move the displacement $e^{-imk_1 \hat{z}}$ to the left of the first chirp, yielding: $e^{-i\xi t_1 \hat{k}^2} e^{-imk_1 \hat{z}} = e^{-imk_1 \hat{z}} e^{-i\xi t_1 (\hat{k}+mk_1)^2}$. Substituting this result back into the original sequence transforms it into:

$$e^{-i\xi t_2 \hat{k}^2} e^{-in\eta k_1 \hat{z}} \left[ e^{-imk_1 \hat{z}} e^{-i\xi t_1 (\hat{k}+mk_1)^2} \right]$$
$$= e^{-i\xi t_2 \hat{k}^2} e^{-i(n\eta+m)k_1 \hat{z}} e^{-i\xi t_1 (\hat{k}+mk_1)^2}, \qquad (S20)$$

where the two adjacent displacement operators have been combined into a single displacement by a total momentum kick of $(n\eta + m)k_1$. We now apply the identity a second time to the remaining pair, $e^{-i\xi t_2 \hat{k}^2} e^{-i(n\eta+m)k_1 \hat{z}}$. This moves the combined displacement fully to the left, giving:

$$e^{-i\xi t_2 \hat{k}^2} e^{-i(n\eta+m)k_1 \hat{z}} = e^{-i(n\eta+m)k_1 \hat{z}} e^{-i\xi t_2 (\hat{k}+(n\eta+m)k_1)^2}. \qquad (S21)$$

The operator sequence for the (n,m) term is thus simplified to:

$$e^{-i(n\eta+m)k_1 \hat{z}} e^{-i\xi t_2 (\hat{k}+(n\eta+m)k_1)^2} e^{-i\xi t_1 (\hat{k}+mk_1)^2}. \qquad (S22)$$

Applying this reordering to every term in the double sum of Eq. (S18), the final state becomes:

$$|\psi_f\rangle = \sum_{n,m} J_n(g_2) J_m(g_1) e^{-in\phi_2} e^{-im\phi_1} e^{-i(n\eta+m)k_1 \hat{z}} e^{-i\xi t_2 (\hat{k}+(n\eta+m)k_1)^2} e^{-i\xi t_1 (\hat{k}+mk_1)^2} |\psi_e^0\rangle. \qquad (S23)$$

This form is highly advantageous, as all displacement operators are now consolidated on the left, and the remaining operators act directly on the initial state in a manner that can be readily evaluated.

## 2.3. Derivation of the Bunching Factor

Substituting the final state from Eq. (S25), $b^{(q)}$ expands into a four-fold sum over sideband indices:

$$b^{(q)} = \sum_{n,\bar{n}} J_{\bar{n}}(g_2) J_n(g_2) e^{i(\bar{n}-n)\phi_2} \sum_{m,\bar{m}} J_{\bar{m}}(g_1) J_m(g_1) e^{i(\bar{m}-m)\phi_1} \langle \psi_e^0 | \hat{B} | \psi_e^0 \rangle, \quad (S24)$$

where the operator $\hat{B}$ encapsulates the combined action of the evolution and the bunching measurement:

$$\hat{B} = e^{i\xi t_1(\hat{k}-\bar{m}k_1)^2} e^{i\xi t_2(\hat{k}-(\bar{n}\eta+\bar{m})k_1)^2} e^{i(\eta\bar{n}+\bar{m})k_1\hat{z}} e^{iqk_1\hat{z}}$$

$$\times e^{-i(\eta n+m)k_1\hat{z}} e^{-i\xi t_2(\hat{k}-(\eta n+m)k_1)^2} e^{-i\xi t_1(\hat{k}-mk_1)^2}. \quad (S25)$$

Our goal is to evaluate the matrix element $\langle \psi_e^0 | \hat{B} | \psi_e^0 \rangle$. This requires simplifying the complex operator $\hat{B}$ and then evaluating its action on the initial state.

The complexity of $\hat{B}$ arises from the non-commutativity of the displacement operators $e^{-ik_1\hat{z}}$ and the chirping operators $e^{-i\xi t\hat{k}^2}$. To simplify, we first combine the three central displacement operators into a single net displacement:

$$e^{i(\eta\bar{n}+\bar{m})k_1\hat{z}} e^{iqk_1\hat{z}} e^{-i(\eta n+m)k_1\hat{z}} = e^{i[q+\eta(\bar{n}-n)+(\bar{m}-m)]k_1\hat{z}}. \quad (S26)$$

We now need to move this consolidated displacement o perator to the left of all chirping operators. This is achieved by repeatedly applying the commutation identity from Eq. (S19), $e^{-i\xi t\hat{k}^2} e^{ik_1\hat{z}} = e^{ik_1\hat{z}} e^{-i\xi t(\hat{k}+k_1)^2}$. Applying this identity systematically moves the displacement leftward, simultaneously shifting the momentum argument in the chirp operators it passes through. After this reordering, the operator $\hat{B}$ simplifies to:

$$\hat{B} = e^{i\xi t_1(\hat{k}-\bar{m}k_1)^2} e^{-i\xi t_1(\hat{k}-(\bar{m}+q+\eta(\bar{n}-n))k_1)^2} e^{i\xi t_2(\hat{k}-(\bar{n}\eta+\bar{m})k_1)^2}$$

$$\times e^{i\xi t_2(\hat{k}-(\bar{n}\eta+\bar{m})k_1)^2} e^{-i\xi t_2(\hat{k}-(\eta\bar{n}+\bar{m}+q)k_1)^2} e^{i(q+\eta(\bar{n}-n)+(\bar{m}-m))k_1\hat{z}}, \quad (S27)$$

We evaluate the matrix element by expanding the initial state in the momentum basis: $|\psi_e^0\rangle = \sum_k c_k|k\rangle$. Thus, $\langle\psi_e^0|\hat{B}|\psi_e^0\rangle = \sum_{k,k'} c_{k'}^* c_k \langle k'|\hat{B}|k\rangle$. The key to evaluating this expression lies in understanding the action of the simplified $\hat{B}$ operator on a momentum eigenstate $|k\rangle$. The operator begins with a displacement, which shifts the momentum state: $e^{i(q+\eta(\bar{n}-n)+(\bar{m}-m))k_1\hat{z}}|k\rangle = |k + (q + \eta(\bar{n} - n) + (\bar{m} - m))k_1\rangle \equiv |k_f\rangle$. The subsequent chirping operators now act on this definite momentum state as complex phase factors. The inner product $\langle k'|k_f\rangle$ enforces momentum conservation via the Dirac delta function $\delta(k' - k_f)$, which collapses the double sum over $k$ and $k'$ by setting $k' = k_f = k + (q + \eta(\bar{n} - n) + (\bar{m} - m))k_1$. Consequently, the matrix element reduces to a single sum over the initial momentum $k$, where the chirping operators contribute specific phase factors evaluated at the shifted momentum arguments, leading to the final form used for further simplification. Thus, we have:

$$\langle\psi_e^0|\hat{B}|\psi_e^0\rangle = \sum_{k',k} \langle k'|k_f\rangle c_{k'}^* c_k \, e^{i\xi t_1\left[(k_f-\bar{m}k_1)^2 - (k_f-(\bar{m}+q+\eta(\bar{n}-n))k_1)^2\right]}$$

$$\times e^{i\xi t_2\left[(k_f-(\bar{n}\eta+\bar{m})k_1)^2 - (k_f-(\eta\bar{n}+\bar{m}+q)k_1)^2\right]}. \quad (S28)$$

where the equation can be further simplified:

$$e^{i\xi t_2\left[(k_f-(\bar{n}\eta+\bar{m})k_1)^2 - (k_f-(\eta\bar{n}+\bar{m}+q)k_1)^2\right]}$$

$$= e^{i\xi t_2\left[(k_f-(\eta\bar{n}+\bar{m})k_1)^2 - (k_f-(\eta\bar{n}+\bar{m})k_1)^2 - (qk_1)^2 + 2(k_f-(\eta\bar{n}+\bar{m})\delta k)(qk_1)\right]}$$

$$= e^{i\xi t_2\left[-(qk_1)^2 + 2(k+(q-\eta n-m)k_1)(qk_1)\right]}$$

$$= e^{i\xi t_2\left[(qk_1)^2 + 2(k-(\eta n+m)k_1)(qk_1)\right]}$$

$$= e^{i2\xi t_2 kqk_1} e^{i\xi t_2\left[q(q-2(\eta n+m))k_1^2\right]} \quad (S29)$$

and

$$e^{i\xi t_1\left[(k_f-\bar{m}k_1)^2-(k_f-(\bar{m}+q+\eta(\bar{n}-n))k_1)^2\right]}$$

$$= e^{i\xi t_1\left[(k_f-\bar{m}k_1)^2-(k_f-\bar{m}k_1-(q+\eta(\bar{n}-n))k_1)^2\right]}$$

$$= e^{i\xi t_1\left[2(k+(q+\eta(\bar{n}-n)+\bar{m}-m)k_1-\bar{m}k_1)(q+\eta(\bar{n}-n))k_1-\left((q+\eta(\bar{n}-n))k_1\right)^2\right]}$$

$$= e^{i\xi t_1\left[2(k+(q+\eta(\bar{n}-n)-m)k_1)(q+\eta(\bar{n}-n))k_1-\left((q+\eta(\bar{n}-n))k_1\right)^2\right]}$$

$$= e^{i\xi t_1\left[2(k-mk_1+(q+\eta(\bar{n}-n))k_1)(q+\eta(\bar{n}-n))k_1-\left((q+\eta(\bar{n}-n))k_1\right)^2\right]}$$

$$= e^{i\xi t_1\left[2(k-mk_1)(q+\eta(\bar{n}-n))k_1+\left((q+\eta(\bar{n}-n))k_1\right)^2\right]}$$

$$= e^{i\xi t_1[2k(q+\eta(\bar{n}-n))k_1]}e^{i\xi t_1\left[-2mk_1(q+\eta(\bar{n}-n))k_1+\left((q+\eta(\bar{n}-n))k_1\right)^2\right]}$$

$$= e^{i\xi t_1[2k(q+\eta(\bar{n}-n))k_1]}e^{i\xi t_1[(q+\eta(\bar{n}-n))(-2m+q+\eta(\bar{n}-n))k_1^2]} \quad (S30)$$

We define $\eta(n-\bar{n}) = \eta q_2$ and $m - \bar{m} = q_1$, so that: $q + \eta(\bar{n}-n) + \bar{m} - m = q - \eta q_2 - q_1$. Again, we simplify the bunching factor

$$b^{(q)} = \sum_{n,\bar{n}} J_{\bar{n}}(g_2) J_n(g_2) e^{-iq_2\phi_2} e^{i\xi t_1[(q-\eta q_2)(-2m+q-\eta q_2)k_1^2]}$$

$$\times \sum_{m,\bar{m}} J_{\bar{m}}(g_1) J_m(g_1) e^{-iq_1\phi_1} e^{i\xi t_2[q(q-2(\eta n+m))k_1^2]}$$

$$\times \sum_{k',k} \langle k'|k_f\rangle c_{k'}^* c_k\, e^{i2\xi k_1(t_1(q-\eta q_2)+t_2 q)k} \quad (S31)$$

Since $\langle k'|k_f\rangle = \delta(k_f - k')$, the last line gives

$$\sum_{k',k} \langle k'|k_f\rangle c_{k'}^* c_k\, e^{i2\xi k_1(t_1(q-\eta q_2)+t_2 q)k} = \sum_k c_{k_f}^* c_k\, e^{i2\xi k_1(t_1(q-\eta q_2)+t_2 q)k} \quad (S32)$$

## 2.4. Derivation under the Large Recoil Condition and Final Expression

The general expression for the bunching factor derived in Section 2.3 remains complex. To obtain the closed-form, interpretable result presented as the central finding in the main text (Eq. 2), we introduce a key physical approximation: the **large recoil condition**. This section details the derivation of Eq. (2) under this condition, explaining the physical origin of each term in the final expression.

The large recoil condition applies when the momentum kick imparted by the laser modulation ($k_1$) is significantly larger than the intrinsic momentum spread of the electron wavepacket ($\sigma_k$). In this regime, the initial state's momentum profile acts as a sharp filter. The product $c_{k_f}^* c_k$, which appears in the sum over $k$, can be approximated as:

$$c_{k_f}^* c_k \approx |c_k|^2 \delta(k_f - k). \tag{S33}$$

The delta function $\delta(k_f - k)$ enforces strict equality between the initial momentum $k$ and the final momentum $k_f = k + (q - \eta q_2 - q_1)k_1$. This directly leads to the constraint: $q - q_1 - \eta q_2 = 0$, or equivalently $q_1 = q - \eta q_2$. This constraint physically means that the net number of photons involved in the entire process must sum to the harmonic order $q$. It collapses the four-fold sum over $m, \bar{m}, n, \bar{n}$ in the general bunching factor into a double sum over $m$ and $n$, with the relationships $\bar{m} = m - q_1 = m - (q - \eta q_2)$ and $\bar{n} = n - q_2$. Summing over the $q_1$ and $k_f$ index, we obtain

$$b^{(q)} = \sum_{n, q_2} J_{n-q_2}(g_2) J_n(g_2) e^{-i q_2 \phi_2} \sum_m J_{m-(q-\eta q_2)}(g_1) J_m(g_1) e^{-i(q-\eta q_2)\phi_1}$$

$$\times e^{i\xi t_1[(q-\eta q_2)(-2m+q-\eta q_2)k_1^2]} e^{i\xi t_2[q(q-2(\eta n+m))k_1^2]}$$

$$\times e^{-2\sigma_k^2(\xi(t_1(q-\eta q_2)+t_2 q)k_1)^2} \tag{S34}$$

where $\bar{n} = n - q_2$, $\bar{m} = m - q_1$. To make further progress, we rearrange the chirping phase factors. Isolating the terms dependent on the summation indices $m$ and $n$ is a crucial step towards applying the next mathematical identity:

$$e^{i\xi t_1[(q-\eta q_2)(-2m+q-\eta q_2)k_1^2]} e^{i\xi t_2[q(q-2(\eta n+m))k_1^2]}$$

$$= e^{i\xi((q-\eta q_2)^2 t_1 + q^2 t_2)k_1^2} e^{i\xi t_2[-2q(\eta n + m)k_1^2]} e^{i\xi t_1[-2m(q-\eta q_2)k_1^2]}$$

$$= e^{i\xi((q-\eta q_2)^2 t_1 + q^2 t_2)k_1^2} e^{-i2\xi(t_2 q + t_1(q-\eta q_2))m k_1^2} e^{-i\xi t_2 2q(\eta n)k_1^2} \quad (S35)$$

Substituting this back, the bunching factor can be rewritten to clearly separate the summations associated with each modulator:

$$b^{(q)} = \sum_{q_1} e^{-iq_2\phi_2} e^{-i(q-\eta q_2)\phi_1} e^{i\xi((q-\eta q_2)^2 t_1 + q^2 t_2)k_1^2}$$

$$\times \sum_n J_{n-q_2}(g_2) J_n(g_2) e^{-i\xi t_2 2q(\eta n)k_1^2}$$

$$\times \sum_m J_{m-(q-\eta q_2)}(g_1) J_m(g_1) e^{-i2\xi(t_2 q + t_1(q-\eta q_2))m k_1^2}$$

$$\times e^{-2\sigma_k^2(\xi(t_1(q-\eta q_2) + t_2 q)k_1)^2} \quad (S36)$$

The summations over $n$ and $m$ are now in a form amenable to evaluation via **Graf's addition theorem** for Bessel functions, which states: $\sum_n J_{n-m}(g) J_n(g) e^{-in\phi} = e^{-im\left(\frac{\pi}{2}+\frac{\phi}{2}\right)} J_m\left[2g \sin\left(\frac{\phi}{2}\right)\right]$, Applying this theorem allows us to perform the infinite sums analytically, yielding compact Bessel function expressions:

$$\sum_n J_{n-q_2}(g_2) J_n(g_2) e^{-i\xi t_2 2q(\eta n)k_1^2} = e^{-iq_2\left(\frac{\pi}{2} + \xi t_2 \eta q k_1^2\right)} J_{q_2}(2g_2 \sin(\xi t_2 \eta q k_1^2)), \quad (S37)$$

$$\sum_m J_{m-(q-\eta q_2)}(g_1) J_m(g_1) e^{-i2\xi(t_2 q + t_1(q-\eta q_2))m k_1^2}$$

$$= e^{-i(q-\eta q_2)\left(\frac{\pi}{2} + \xi(q t_2 + (q-\eta q_2)t_1)k_1^2\right)} J_{q-\eta q_2}(2g_1 \sin(\xi(q t_2 + (q-\eta q_2)t_1)k_1^2)). \quad (S38)$$

Substituting Eqs. (S37) and (S38) back into Eq. (S36) and consolidating all the complex phase factors, we arrive at the final expression for the bunching factor amplitude. Its modulus, which gives the spectral amplitude of the q-th harmonic, is the central result presented in the main text:

$$b^{(q)} = e^{i\xi q^2 t_2 k_1^2} \sum_{q_2} e^{-iq_2\phi_2} e^{-i(q-\eta q_2)\phi_1} e^{i\xi((q-\eta q_2)^2 t_1)k_1^2}$$

$$\times e^{-iq_2\left(\frac{\pi}{2}\right)} e^{-iq_2(\xi t_2 \eta q k_1^2)} J_{q_2}(2g_2 \sin(\xi t_2 \eta q k_1^2))$$

$$\times e^{-i(q-\eta q_2)\left(\frac{\pi}{2}+\xi(qt_2+(q-\eta q_2)t_1)k_1^2\right)} J_{q-\eta q_2}(2g_1 \sin(\xi(qt_2 + (q-\eta q_2)t_1)k_1^2))$$

$$\times e^{i2\xi k_1(t_1(q-\eta q_2)+t_2 q)k}$$

$$= \sum_{q_2} e^{-iq\left(\frac{\pi}{2}+\phi_1\right)} e^{iq_2\left(\eta\phi_1 - \phi_2 + \frac{\pi}{2}(\eta-1)\right)}$$

$$\times J_{q-\eta q_2}(2g_1 \sin(\xi(qt_2 + (q-\eta q_2)t_1)k_1^2))$$

$$\times J_{q_2}(2g_2 \sin(\xi t_2 \eta q k_1^2))$$

$$\times e^{-2\sigma_k^2(\xi(t_1(q-\eta q_2)+t_2 q)k_1)^2} \tag{S39}$$

Thus, we have:

$$|b^{(q)}| = \left| \sum_{q_2} e^{iq_2\Theta} J_{q-\eta q_2}(2g_1 \sin(\xi((q-\eta q_2)t_1 + qt_2)k_1^2)) \right.$$

$$\left. \times J_{q_2}(2g_2 \sin(\xi t_2 \eta q k_1^2)) e^{-2((q\xi(t_1+t_2)-\eta q_2 \xi t_1)\sigma_k k_1)^2} \right| \tag{S40}$$

This final form elegantly encapsulates the entire physics of the quantum echo process. The summation index $q_2$ runs over all possible quantum pathways, representing the net number of photons absorbed or emitted in the second modulator. The phase $\Theta = \eta\phi_1 - \phi_2 + \frac{\pi}{2}(\eta - 1)$ is the effective phase difference controlling interference between these pathways. The Bessel functions $J_{q-\eta q_2}$ and $J_{q_2}$ encode the probability amplitudes for the multiphoton transitions in the first and second modulators, respectively. The phases $M_1^{(q_2)} = \xi((q - \eta q_2)t_1 + qt_2)k_1^2$ and $M_2 = \xi t_2 \eta q k_1^2$ represent the dispersive phases accumulated during free propagation, with $M_1^{(q_2)}$ being pathway-dependent. Finally, the envelope function $U^{(q_2)} = \exp\left(-2((q\xi(t_1 + t_2) - \eta q_2 \xi t_1)\sigma_k k_1)^2\right)$ accounts for the suppression of coherence due to the finite momentum spread of the electron wavepacket. This derivation thereby establishes a direct and rigorous connection between the fundamental operator-based model and the practical, interpretable formula used to describe and engineer quantum echo high-harmonic generation.

## 3. Analysis of Special Cases

To validate the internal consistency of our general quantum echo model and to elucidate the distinct physical roles of each component in the beamline, we examine key limiting cases of the central result in Eq. (2). These analytical simplifications serve as critical checks, confirming that our formalism reduces to known results in appropriate limits and highlighting the unique physical consequences of the full quantum echo sequence.

### 3.1. Single-Modulation Limit

Setting the modulation strength of the second laser to zero ($g_2 = 0$) allows us to recover the established physics of standard PINEM-based high-harmonic generation. In this limit, the Bessel function associated with the second modulator simplifies to $J_{q_2}(0) = \delta_{q_2,0}$, which collapses the summation over the quantum pathway index $q_2$ in Eq. (2) to a single term. Substituting $q_2 = 0$ into the general expression yields the bunching factor for a single modulation:

$$|b^{(q)}| = |J_q(2g_1 \sin(\xi q(t_1 + t_2)k_1^2))| \exp\left(-2\sigma_k^2 (\xi q(t_1 + t_2)k_1)^2\right). \quad (S41)$$

Introducing the total effective chirp duration $\tau = \xi(t_1 + t_2)$, this simplifies to the compact and familiar form:

$$|b^{(q)}| = |J_q(2g_1 \sin(q\tau k_1^2))| \exp\left(-2(q\tau \sigma_k k_1)^2\right). \quad (S42)$$

This result is identical to the classical single-modulation PINEM-HHG result, confirming that our general quantum echo formalism correctly reduces to the established single-modulator physics.

## 3.2. Two-Color Modulation without Chirp 1

We next investigate the critical role of the chirp sections by considering the case where the first chirp is removed ($t_1 = 0$). This configuration eliminates the pathway-dependent phase accumulation from the first free-propagation stage. Consequently, the two modulations no longer act as independent, time-separated events but rather coalesce into an effective, simultaneous two-color interaction.

Setting $t_1 = 0$ in the general expression (Eq. 2) simplifies the relevant terms, and the bunching factor becomes:

$$|b^{(q)}| = |\sum_{q_2} J_{q-\eta q_2}(2g_1 \sin(q\tau_2 k_1^2)) J_{q_2}(2g_2 \sin(\eta q\tau_2 k_1^2))| \exp(-2(q\tau_2 \sigma_k k_1)^2) \quad (S43)$$

where $\tau_2 = \xi t_2$. The summation over $q_2$ is now in a standard form that can be evaluated analytically using the Bessel function addition theorem: $\sum_{m=-\infty}^{\infty} J_{\nu-m}(A) J_m(B) = J_\nu(A+B)$, where $\nu$ is an integer. Identifying $\nu = q$, $m = q_2$, $A = 2g_1 \sin(q\tau_2 k_1^2)$, and $B = 2g_2 \sin(\eta q\tau_2 k_1^2)$, the summation simplifies directly to a single Bessel function: $\sum_{q_2} J_{q-q_2}(A) J_{q_2}(B) = J_q(A+B)$.

The final expression for the bunching factor thus reduces to:
$$|b^{(q)}| = \exp(-2(q\tau_2 \sigma_k k_1)^2) |J_q(2g_1 \sin(q\tau_2 k_1^2) + 2g_2 \sin(\eta q\tau_2 k_1^2))|. \quad (S44)$$

This result describes a two-color PINEM-HHG process where the harmonic amplitude is governed by a single Bessel function whose argument is the sum of the contributions from both modulators. The absence of Chirp 1 removes the ability to resolve and selectively interfere individual quantum pathways ($q_2$), causing all pathways to add constructively as if the modulations occurred concurrently. This starkly underscores that Chirp 1 is not a passive drift section but an active phase-processing component essential for enabling the pathway-selective interference that defines the quantum echo mechanism. The destructive interference between pathways, which is key to harmonic selectivity in the full QEEHG scheme, is lost in this limit.

**4. Momentum-Space Analysis of the Bunching Factor**

To develop an intuitive understanding of the harmonic generation process and to connect with the computational analysis presented in the main text (Figs. 2e, 2f), it is insightful to transition from the position-space representation of the electron wavefunction to its momentum-space representation. This approach provides a direct view of how different momentum components interfere to produce the final harmonic signal.

The q-th harmonic bunching factor is fundamentally defined as the quantum mechanical expectation value $b^{(q)} = \langle \psi_f | \hat{b}^q | \psi_f \rangle$, where $\hat{b} = e^{ik_1 \hat{z}}$ is the bunching operator. While this definition is physically transparent, a deeper understanding of the harmonic generation process emerges when we transition to the momentum-space representation. This approach reveals how different momentum components of the final electron wavefunction interfere to produce the coherent radiation at harmonic frequency $q\omega$, providing the foundation for the phase-resolved analyses presented in Figs. 2e and 2f of the main text.

We begin by inserting the identity operator expressed in the momentum basis, $I = \int dp |p\rangle\langle p|$, into the definition of the bunching factor:

$$b^{(q)} = \langle \psi_f | \hat{b}^q | \psi_f \rangle$$

$$= \int dp \int dp' \langle \psi_f | p' \rangle \langle p' | e^{iqk_1 \hat{z}} | p \rangle \langle p | \psi_f \rangle. \quad (S45)$$

Recognizing that $\langle p | \psi_f \rangle = \psi_f(p)$ is the momentum-space wavefunction and that the bunching operator acts as a displacement in momentum space, $\langle p' | e^{iqk_1 \hat{z}} | p \rangle = \delta(p' - (p + qk_1))$, the expression simplifies to a single integral:

$$b^{(q)} = \int dp \, \psi_f^*(p + qk_1) \, \psi_f(p). \quad (S46)$$

This momentum-space formulation offers a powerful interpretation: the bunching factor $b^{(q)}$ is the overlap integral between the electron's final momentum wavefunction and a version of itself shifted by the harmonic recoil momentum $qk_1$. Physically, it quantifies the coherence between momentum states separated by $qk_1$, which is the fundamental condition for the constructive interference that leads to the emission of radiation at the harmonic frequency $q\omega$.

**4.1 Phase-Resolved Momentum-Space Decomposition**

While Eq. (S46) provides the total complex amplitude of the q-th harmonic, our objective is to understand *how* different regions of the momentum distribution contribute to this net signal. This deeper analysis is crucial for interpreting the quantum pathway interference demonstrated in Figs. 2e and 2f of the main text.

In numerical simulations, we evaluate a discretized version of this integral:

$$b^{(q)} = \sum_p C(p), \text{where } C(p) = \psi_f^*(p + qk_1)\,\psi_f(p)\Delta p. \qquad (S47)$$

Here, each $C(p)$ is a complex quantity representing the contribution from a specific momentum component $p$. The modulus $|b^{(q)}|$ determines the observable harmonic strength, while its global phase, $\phi = \arg(b^{(q)})$, encodes the precise timing of the emission.

To deconstruct which momentum components interfere constructively or destructively to form the total signal, we project each complex contribution $C(p)$ onto the phase direction of the total harmonic field $b^{(q)}$. This is achieved by calculating the real part of each term after a phase rotation:

$$C_{\text{proj}}(p) = \text{Re}\left[C(p)e^{-i\phi}\right], \text{with } \phi = \arg(b^{(q)}). \qquad (S48)$$

**Interpretation:**

- $C_{\text{proj}}(p) > 0$ : The momentum component at $p$ interferes **constructively**, enhancing the harmonic yield.

- $C_{\text{proj}}(p) < 0$ : The momentum component at $p$ interferes **destructively**, suppressing the harmonic yield.

This phase-resolved decomposition, $C_{\text{proj}}(p)$, is the analytical tool used to generate the harmonic-channel-resolved analyses in the main text. It unambiguously isolates the phase-sensitive contributions from across the electron's momentum profile, visually confirming how the quantum echo process actively orchestrates constructive interference in desired harmonic channels while suppressing competing pathways.

## 5. Numerical Framework and Optimization Algorithm

The theoretical model derived in Section 2 provides an analytical foundation for understanding quantum echo high-harmonic generation. However, for optimization across the multi-dimensional parameter space, numerical simulation becomes essential. This section details our computational approach and formalizes the optimization challenge central to realizing the QEEHG scheme.

### 5.1. Quantum Wavepacket Evolution Algorithm

The quantum evolution of the electron wavepacket through the QEEHG beamline is simulated by sequentially applying the modulation and free-propagation operators. The key numerical consideration is that the laser modulation operator $\widehat{M}(g) = \exp(-ig\sin(k_1\hat{z} + \phi))$ is naturally applied in position space, while the chirping operator $\widehat{U}_{\text{free}}(t) = \exp(-i\xi t\hat{k}^2)$ is most efficiently applied in momentum space. This necessitates repeated transformations between the position and momentum representations using the Fast Fourier Transform (FFT) algorithm. The complete numerical procedure, implemented as the function ElectronModu, is outlined below:

| Step | Operation | Space |
|------|-----------|-------|
| 1 | `Initialize: ψ_z0 ← Gaussian(z, σ_z)`<br>Initialize electron as Gaussian wavepacket | Position |
| 2 | `First Modulation: ψ_z(m1) = exp(-i g₁ sin(k₁z + φ₁)) × ψ_z0`<br>Apply first PINEM modulation | Position |
| 3 | `FFT: ψ_p(m1) = FFT[ψ_z(m1)]`<br>Transform to momentum space | Momentum |
| 4 | `First Propagation: ψ_p(b1) = exp(-i ξ t₁ k²) × ψ_p(m1)`<br>Apply dispersive chirp (t₁ = d₁/v₀) | Momentum |
| 5 | `IFFT: ψ_z(b1) = IFFT[ψ_p(b1)]`<br>Transform to position space | Position |
| 6 | `Second Modulation: ψ_z(m2) = exp(-i g₂ sin(ηk₁z + φ₂)) × ψ_z(b1)`<br>Apply second PINEM modulation | Position |
| 7 | `FFT: ψ_p(m2) = FFT[ψ_z(m2)]`<br>Transform to momentum space | Momentum |
| 8 | `Second Propagation: ψ_p(b2) = exp(-i ξ t₂ k²) × ψ_p(m2)`<br>Apply dispersive chirp (t₂ = d₂/v₀)<br>*Final state in momentum space: ψ_f = ψ_p(b2)* | Momentum |
| 9 | `IFFT (Optional): ψ_z(b2) = IFFT[ψ_p(b2)]`<br>Final transformation for analysis<br>*Final state in position space: ψ_f = ψ_z(b2)* | Position |
| 10 | `Compute: b^(q) = ⟨ψ_f | e^(i q k₁ ẑ) | ψ_f⟩`<br>Calculate harmonic amplitude using final state ψ_f<br>*ψ_f is defined in step 8 (momentum space) or step 9 (position space)* | Result |

Fig. S1: The snapshot of the QEEHG algorithm.

This algorithm faithfully reproduces the quantum echo sequence described by Eq. (1) in the main text and serves as the computational engine for all numerical results presented.

5.2 Optimization Framework for QEEHG

The central challenge in practical QEEHG implementation is identifying the optimal set of experimental parameters that maximize harmonic generation at a desired frequency. Formally, the q-th harmonic spectrum $|b^{(q)}|^2$ is a scalar function of six independent control parameters: $\mathbf{\Pi} = \{|g_1|, \phi_1, d_1, |g_2|, \phi_2, d_2\}$, comprising the modulation strengths, optical phases, and drift lengths. The QEEHG optimization problem is defined as finding the parameter set $\boldsymbol{\Pi}^*$ that maximizes the bunching factor for a designated target harmonic $q_d$: $\Pi^*_{\eta,\lambda_1} = \arg \max_{\Pi_{\eta,\lambda_1}} |b^{(q_d)}|$.

The optimization of QEEHG parameters presents unique challenges due to the vastly different scales and physical meanings of the six control parameters. Modulation strengths ($g_1$, $g_2$) are dimensionless and typically range from 0 to ~100, laser phases ($\varphi_1$, $\varphi_2$) are in radians (0 to $2\pi$), while drift lengths ($d_1$, $d_2$) are in micrometers and can span several orders of magnitude. This parameter heterogeneity necessitates careful scaling strategies in the optimization algorithm.

To efficiently navigate the high-dimensional parameter space, we employ a gradient-based optimization approach. The optimization employs a scaled gradient descent approach that accounts for parameter heterogeneity. The scaling gradient method ensures that all parameters contribute appropriately to the optimization direction, thereby achieving more efficient and reliable convergence.